\documentclass[prb,twocolumn,superscriptaddress,showpacs,preprintnumbers,amsmath,amssymb]{revtex4}

\usepackage{amsmath,amssymb}
\usepackage[dvipsnames]{xcolor}

\usepackage{epsfig}
\usepackage{dcolumn}
\usepackage{bm}
\usepackage{soul}

\definecolor{darkgreen}{rgb}{0,0.6,0}

\begin{document}

\title{
Single-Parameter Scaling and Maximum Entropy inside Disordered One-Dimensional Systems: Theory and Experiment.
}

\author{Xiaojun Cheng}
\thanks{These three authors contributed equally}
\address{Department of Physics, Queens College and the Graduate Center of the City University of New York, Flushing, NY, 11367 USA}

\author{Xujun Ma}
\thanks{These three authors contributed equally}
\address{Department of Physics, Queens College and the Graduate Center of the City University of New York, Flushing, NY, 11367 USA}

\author{Miztli Y\'epez}
\thanks{These three authors contributed equally}
\affiliation{Departamento de F\'isica, Universidad Aut\'onoma Metropolitana-Iztapalapa, A. P. 55-534, 09340 M\'exico D. F., Mexico}

\author{Azriel Z. Genack}
\address{Department of Physics, Queens College and the Graduate Center of the City University of New York, Flushing, NY, 11367 USA}

\author{Pier A. Mello}
\address{Instituto de F\'isica, Universidad Nacional Aut\'onoma de M\'exico, Ap. Postal 20-364, 01000 M\'exico, D. F., Mexico}

́
\begin{abstract} 

The single-parameter scaling hypothesis relating the average and variance of the logarithm of the conductance is a pillar of the theory of electronic transport.
We use a maximum-entropy ansatz to explore the logarithm of the energy density, 
$\ln {\cal W}(x)$, at a depth $x$ into a random one-dimensional system.
Single-parameter scaling would be the special
case in which $x=L$ (the system length).
We find the result, confirmed in microwave measurements and computer simulations, that the average of $\ln {\cal W}(x)$ is independent of $L$ and equal to $-x/\ell$, with $\ell$ the mean free path. 
At the beginning of the sample, ${\rm var}[\ln {\cal W}(x)]$ 
rises linearly with $x$
and is also independent of $L$,
with a sublinear increase near the sample output.
At $x=L$ we find a correction to the value
of ${\rm var}[\ln T]$
predicted by single-parameter scaling.

\end{abstract}



\maketitle

Studies of electronic transport have focused on the
scaling of the conductance. 
As a result of the equivalence of the electronic conductance expressed in units of the quantum of conductance and the transmittance of classical waves, many of the predictions of mesoscopic physics and localization theory apply equally to the transport of quantum and classical waves 
[\onlinecite{cheng1a,cheng1b,cheng1c,cheng1d,cheng1e,cheng1f,cheng1g,cheng1h}]. Classical waves are temporally coherent in random static samples so that mesoscopic aspects of propagation are manifest even in macroscopic samples at room temperature and measurements can be carried out in ensembles of statistically equivalent samples 
[\onlinecite{cheng1f,cheng1g}]. 
In addition to studies of conductance and transmission, the 
statistics of transport inside random systems has been studied for many years 
[\onlinecite{gazaryan,kohler_papanicolau,neupane_yamilov,van_tiggelen_et_al_2000-2006,tian_et_al_2010}]. 
Interest in waves in the interior of random samples has intensified recently because of the possibility of exploiting measurements of the transmission matrix 
[\onlinecite{2008b,2010c}] to control waves transmitted through and within the interior 
[\onlinecite{cheng2a,cheng2b,cheng2c,cheng2d,cheng2e,cheng2f,cheng2g,cheng2h}]
by preparing the incident wave in specific transmission eigenchannels [\onlinecite{cheng_et_al_2014}].

A key assumption in the theory of wave transport is that
the scaling and statistics of the transport depend upon
a single parameter. The single parameter scaling (SPS)
hypothesis holds that, in the localized regime, the distribution of the logarithm
of the conductance or transmittance is a Gaussian with
variance equal to twice 
the magnitude of its average value [\onlinecite{beenakker97}], 
${\rm var}(\ln T)=- 2\langle \ln T \rangle$. 
Here, $\langle \cdots \rangle$ indicates
the average over statistically equivalent samples. 
SPS has aided in understanding the statistics
of the logarithm of transmission.
However, the possibility of finding the 
expectation value of the logarithm of the energy density in the interior of random media and relating it to the 
corresponding variance  has not been considered.
Since SPS would be a special case of such a general treatment, in which $x \to L$,
this allows us to test SPS.
Aside from its fundamental importance, this can provide a guide to effective strategies for imaging and energy deposition.

In this Rapid Communication, we study the statistics of particle and energy density in the interior of random samples
applying a maximum-entropy approach (MEA) [\onlinecite{mello-kumar}] 
to random-matrix theory. 
We find the simple result
$\left\langle \ln {\cal W}(x) \right\rangle =-x/\ell$,
where ${\cal W}(x)$ is the energy density at depth $x$ normalized so that its value for $x=L$ is $T$, 
and $\ell$ is the elastic mean free path. 
Though $\langle{\cal W}(x)\rangle$ at depth $x$ increases as the sample length increases, since a larger fraction of the wave energy that reaches $x$ returns to $x$ in samples of larger $L$, nonetheless, 
$\left\langle \ln {\cal W}(x) \right\rangle$ is unchanged as $L$ increases.
In the localized regime, the probability distribution function (PDF) of $\ln {\cal W}(x)$ is Gaussian away from the sample input, 
its variance increasing linearly with depth $x$ from the sample input boundary until it begins to fall near the output surface. 
In the regime 
where
the variance of  $\ln {\cal W}(x)$ increases linearly with $x$, it is 
also 
independent of $L$.
These results 
are confirmed in microwave measurements and computer simulations in random single-mode waveguides.

The MEA of Ref. [\onlinecite{mello-kumar}]
is a random-matrix theory which leads to a Fokker-Planck equation,
known as the Dorokhov-Mello-Pereyra-Kumar (DMPK) equation
[\onlinecite{dorokhov_82,mpk}],
governing the ``evolution" with sample length $L$ of the 
PDF $p_{L}(M)$ of the system transfer matrix $M$.
The multiplicative matrix $M$ is the random matrix of this theory.
In the MEA the disordered system is assumed to contain a large number of 
weak scatterers.
An {\em ansatz} is proposed for the PDF of the transfer matrix for a thin piece of material, a ``building block", 
which contains the physical information relevant to the problem: 
the Shannon entropy of 
$p(M)$ for a building block is maximized constrained by 
normalization and a given $\ell$.
The PDF for the full system is then 
constructed by successive convolutions.
In this dense-weak-scattering limit the MEA is expected to give results insensitive to microscopic details.
This is a ``local approach", in contrast with the so-called 
``global approach" 
[\onlinecite{stone_mello_et_al_1991}].


The DMPK equation was developed for $N$ (propagating) modes.
For one dimension (1D), the DMPK equation [\onlinecite{mpk}] reduces to Melnikov's equation [\onlinecite{cheng1d}].
The study of the statistical properties of the intensity profile inside random 1D samples 
using Melnikov's equation was initiated in 
Ref. [\onlinecite{mello_genack_et_al_2015}]: 
the expectation value of the energy density, ${\cal W}(x)$, was obtained and compared successfully with computer simulations
(see also Ref. [\onlinecite{freilikher_2003}]).
In the present paper we extend this analysis to investigate 
the statistics of the self-averaging quantity $\ln {\cal W}(x)$,
not contemplated in Ref. [\onlinecite{mello_genack_et_al_2015}].
These studies may provide a path for the extension to 
quasi-1D disordered systems supporting more than a single open channel. 

Consider the scattering in a 1D random distribution of scatterers, as illustrated in 
the top part of Fig. \ref{<lnW(x)>, var_lnW(x) theo. vs simuls.}.
This situation may arise: 
i) in a quantum-mechanical (QM) problem describing electronic scattering in a disordered conductor, or
ii) in the problem of an electromagnetic (EM) wave in a disordered waveguide supporting a single transverse mode, or of a plane wave impinging upon a random layered medium.
The amplitudes of the incident, transmitted, and reflected waves are also indicated.
We imagine opening a gap, which is small compared to the wavelength at the point $x$ inside the sample, as shown in
Fig. \ref{<lnW(x)>, var_lnW(x) theo. vs simuls.}, 
where the amplitudes of the waves travelling to the right and left are shown
(continuity of the wavefunction and its derivative are imposed).
\begin{figure}[t]
\centerline{
\includegraphics[width=8.5cm,height=5.0cm]
{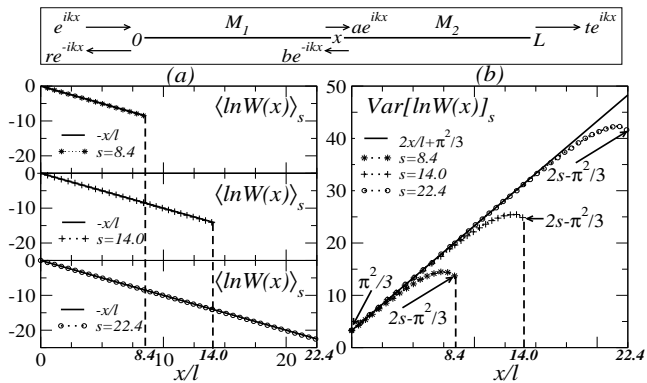}
}
\caption{
Top panel: the scattering problem for the 1D disordered waveguide of length $L$ described in the text.
Lower panels:
theoretical results (full lines) and computer simulations 
(various symbols and dotted lines) for the profiles 
$\left\langle  \ln {\cal W}(x) \right\rangle_{s}$,
(panel (a)), and
${\rm var} [\ln {\cal W}(x)]_s$,
(panel (b)), 
as functions of $x/\ell$, for three values of $s=L/\ell$.
For the variance, 
Eq. (12) 
was complemented with 
Eq. (13) 
when $x=L$  
(see the arrows):
their combination accounts for the ``bending" 
shown by the simulation.
Agreement is excellent.
Simulations consist of $10^5$ realizations,
with $kd=0.1$ and $k\ell=178$.
}
\label{<lnW(x)>, var_lnW(x) theo. vs simuls.}
\end{figure}

Inside the gap, the intensity is
\begin{equation}
{\cal W}(x) 
= |a e^{ikx} + be^{-ikx}|^2.
\label{W(x) 0}
\end{equation}
Writing
the transfer matrices 
of the two segments as
\begin{equation}
M_i = 
\left[
\begin{array}{cc}
\alpha_i   & \beta_i \\
\beta_i^{*} & \alpha_i^{*}
\end{array}
\right], 
\;\;\;\;\; 
i=1,2,
\label{M_i}
\end{equation}
with 
$|\alpha_i|^2 - |\beta_i|^2 =1$, 
we satisfy the requirements of time-reversal invariance and flux conservation.
When no index $i$ is employed, 
we refer to the system as a whole.
The intensity 
of Eq. ({\ref{W(x) 0}), denoted here as ${\cal W}(x; M_1, M_2)$,
is [\onlinecite{mello_genack_et_al_2015}]
\begin{equation}
{\cal W}(x; M_1, M_2)
= \frac{|\alpha_2^{*}e^{ikx} - \beta_2^{*}e^{-ikx}|^2}
{|\alpha_2 \alpha_1 + \beta_2 \beta_1^{*}|^2}
\equiv T F_x(M_2) \; , 
\label{W(x)}
\end{equation}
where $k$ denotes the wavenumber and $T(M_1,M_2)$ the transmission coefficient of the full sample.
In the polar decomposition defined in Ref. [\onlinecite{mello-kumar}], 
the transfer matrices $M_i$ can be written in terms of ``radial parameters"
$\lambda_i \ge 0$ [$T_i=1/(1+\lambda_i)$] 
and two phases, $\theta_i$ and $\mu_i$,
as $\alpha_i = \sqrt{1+ \lambda_i}\exp(i \theta_i),
\;\;\beta_i= \sqrt{\lambda_i} \exp(i(2\mu_i -\theta_i))$.
The function $F_x(M_2)$ in Eq. (\ref{W(x)}) is then
\begin{equation}
F_x(M_2)
= A(\lambda_2) -B(\lambda_2)
\cos2(\mu_2-\theta_2 + kx) \; ,
\label{Fx(M2)}  
\end{equation}
with $A(\lambda_2)=1+2\lambda_2$ and $B(\lambda_2)=2\sqrt{\lambda_2(1+\lambda_2)}$.

The above expressions refer to a single configuration of disorder.
Assuming the disorder is uncorrelated, 
quantities associated with the two sections 
of the sample are statistically independent of one another. 
The expectation value over an ensemble of configurations of a function 
$f({\cal W}(x))$ can be computed using the PDF of the transfer matrices for the two sections, $p_{x}(M_1)$ and $p_{L-x}(M_2)$.
For samples of length $L$,
Melnikov's diffusion equation governs the evolution with 
$s=L/\ell$ 
of the marginal PDF $ w_s(\lambda)$
of the radial parameter $\lambda$ as  
\begin{equation}
\frac{\partial w_s(\lambda)}{\partial s}
= \frac{\partial}{\partial \lambda}
\left[\lambda (1+ \lambda) 
\frac{\partial w_s(\lambda)}{\partial \lambda}
\right] \; .
\label{melnikov}
\end{equation}
Equation (\ref{melnikov}) is solved with the initial condition
$
w_{s=0}(\lambda) = \delta_{+}(\lambda),
$
where $\delta_{+}(\lambda)$ is a one-sided delta function.
In what follows, the statistics of each one of the radial parameters $\lambda_1$, $\lambda_2$ of the two statistically independent sections of the wire will be described by 
Eq. (\ref{melnikov}):
for the left segment,
$s$ will be replaced by $s_1=x/\ell$, and
for the right segment, by $s_2=(L-x)/\ell$.

From Eq. (\ref{W(x)}), we find for the ensemble average
\begin{equation}
\langle \ln {\cal W}(x) \rangle_s
= \langle \ln T \rangle_{s}
+\langle \ln F_x(M_2) \rangle_{s_2} \; .
\label{<ln W> 1}
\end{equation}
The first term is given by the well-known expression
\begin{equation}
\langle \ln T \rangle_{s}
\equiv \int_0^{\infty}\ln T w_s(\lambda) d\lambda.
= -\frac{L}{\ell} \; .
\label{<ln T>}
\end{equation}
From Eq. (\ref{Fx(M2)}), the second term can be written as
\begin{eqnarray}
&&
\langle \ln F_x(M_2) \rangle_{s_2} 
  = \int_0^{\infty} d\lambda_2 \int_0^{2\pi} d\theta_2
w_{s_2}(\lambda_2, \theta_2) 
\int_{0}^{2\pi} \frac{d\mu_2}{2\pi} 
\nonumber \\
&& \times 
\ln \big[A(\lambda_2) - B(\lambda_2)\cos(2(\mu_2-\theta_2 + kx))\big] =  s_2 ,   
\label{<ln W> 2b}   
\end{eqnarray}
where we used Eq. (4.224.9) of Ref. [\onlinecite{gradshteyn}] to evaluate the angular integral in Eq. (\ref{<ln W> 2b}).
The final result is
\begin{equation}
\left\langle  \ln {\cal W}(x) \right\rangle_s
= - \frac{x}{\ell}.
\label{<ln W>}
\end{equation}
Notice that {\em the $L$ dependence has dropped out} from this result.
A simple demonstration of this independence for $x=0$ is given in                            
the supplemental material (SM) presented in 
Ref. [\onlinecite{SM_1}];
it uses the statistics of the reflection amplitude $r$ of Ref. [\onlinecite{yepez-jjs}].
For $x=L$, Eq. (\ref{<ln W>}) reduces to Eq. (\ref{<ln T>})
for the full sample.  
We may alternatively use the identity (26) of
Ref. [\onlinecite{Anderson_1980}], to show 
the independence of the result on $L$.

From Eq. (\ref{W(x)}), the second moment of $\ln {\cal W}(x)$ is
\begin{eqnarray}
\left\langle \left[ \ln {\cal W}(x) \right]^2  \right\rangle_s
= && \left\langle (\ln T)^2 \right\rangle_{s}
+\Big\langle [\ln F_x(M_2)]^2 \Big\rangle_{s_2} 
\nonumber \\
&& + 2 \Big\langle 
(\ln T) \left[ \ln F_x(M_2)\right] 
\Big\rangle_{s,s_2}  \; .
\label{<(ln W)2>}
\end{eqnarray}
Although we have not succeeded in computing the 
three terms
in Eq. (\ref{<(ln W)2>}) for 
arbitrary 
$s=L/\ell$, we have found approximate expressions for the case when the wave in the right segment is localized:
$s, s_2 \gg 1$.
We obtain
\begin{subequations}
\begin{eqnarray}
&& 
\left\langle (\ln T)^2 \right\rangle_{s}
= s^2 + 2s -2C + \omega_1(s), 
\label{<(lnT)2>}  \\
&& 
\Big\langle [\ln F_x(M_2)]^2 \Big\rangle_{s_2}
= s_2^2 + 2 s_2 
+( \frac{\pi^2}{3} -2C) 
+ \omega_2(s_2) ,
\\
\label{<(ln W)2> 2}  
&& 2\Big\langle 
(\ln T) \left[ \ln F_x(M_2)\right] 
\Big\rangle_{s, s_2}
= -2s_2 s -4s_2 +4C+ \omega_3(s_2),
\nonumber  \\
\label{<(ln W)2> 3}
\end{eqnarray}
\label{<(ln W)2> 123}
\end{subequations}
\noindent
where [\onlinecite{marlos_et_al}]
$
C \equiv \int_0^{\infty}\langle T \rangle_s ds 
= \pi^2 /6.
$ 
Collecting terms and using Eq. (\ref{<ln W>}), we find for the variance of $\ln {\cal W}(x)$
\begin{equation}
{\rm var}\left[ \ln {\cal W}(x) \right]_s
= 2 \frac{x}{\ell} + \pi^2 /3 + \omega_{4}(s_2) ,  \;\;\;\;   s, s_2 \gg 1 .
\label{var(ln W)}
\end{equation}
In Eqs. (\ref{<(ln W)2> 123}) and (\ref{var(ln W)}), $\omega_i(s)$ are functions that tend to 0 as $s \to \infty$
(e.g., $\omega_1=\int_0^{s}\langle T \rangle_s ds-  \pi^2 /6$).
For $x=L$ we cannot apply the above result, Eq. (\ref{var(ln W)}), since this would violate the condition $s_2 \gg 1$.
Since for $x=L$, ${\cal W}(L)=T$, one finds, from Eqs. (\ref{<(lnT)2>}) and (\ref{<ln T>})
\begin{equation}
({\rm var}(\ln T))_{s}
= 2\frac{L}{\ell} -\pi^2 /3 + \omega_1(s),
\hspace{5mm}s \gg 1 \; .
\label{var(ln T)}
\end{equation}
To leading order in $s \gg 1$, Eq. (\ref{var(ln T)}) can be approximated by its first term, which represents the well-known result that the variance of the logarithm of the transmission scales as twice (the absolute value of) its expectation value; 
in addition, $\ln T$ has a normal probability distribution 
[\onlinecite{mello86,beenakker97}], with 
$\langle \ln T \rangle = -s$, $({\rm var}(\ln T))_{s} = 2s$. 
The next term in Eq. (\ref{var(ln T)}), i.e., $-\pi^2 /3$, 
represents a correction to $({\rm var}(\ln T))_{s}$ of order
$s^0$.
To the best of our knowledge, this correction has not been reported before:
earlier studies were restricted to lowest order in $s$;
this correction may not be negligible if $s$ is not large 
(see, e.g., Fig. \ref{<lnW(x)>, var_lnW(x) theo. vs simuls.}(b), 
explained below).

To check results, we have carried out computer simulations
of random waveguides 
supporting a single propagating mode.
These simulations can be applied to both the QM and EM cases:
i) in the QM case, the disordered potential is a random function of position;
we chose sequences of equidistant barriers 
(idealized as delta-function potentials), with separation $d$
small compared with the wavelength; 
ii) in the EM case, it is the index of refraction $n$ 
appearing in the Helmholtz equation which is a similar random function of position.

The profiles of $\left\langle  \ln {\cal W}(x) \right\rangle_s$
and ${\rm var} [\ln {\cal W}(x)]_s$,
Eqs. (\ref{<ln W>}) and (\ref{var(ln W)}),
are shown, as functions of 
$x/\ell$, for three values of $s$, in panels (a) and (b) of 
Fig. \ref{<lnW(x)>, var_lnW(x) theo. vs simuls.}.
The results in (a) show that 
$\left\langle  \ln {\cal W}(x) \right\rangle_s$
is insensitive to $s=L/\ell$, while the results in (b) 
show that ${\rm var} [\ln {\cal W}(x)]_s$
is insensitive in the linear regime.
Simulations are in excellent agreement with theoretical results.   
From Eq. (\ref{var(ln W)}), the theoretical variance for $x=0$ has the 
{\em universal} value $\pi^2/3$ in the localized regime, which agrees with simulation.
A simple derivation of this result is given in the SM presented in
Ref. [\onlinecite{SM_1}].
The first moment and variance of $\ln {\cal W}(x)$
are shown as functions of $s$ for fixed values of $x/L$
in panels (a) and (b), respectively, of Fig. 1 
of the SM of Ref. [\onlinecite{SM_1}]:
they continue to the interior of the sample
the results at $x=L$  
for $\langle \ln T \rangle_s$ and 
${\rm var}[\ln T]_s$.
\begin{figure}[t]
\centerline{
\includegraphics[width=8cm,height=5cm]{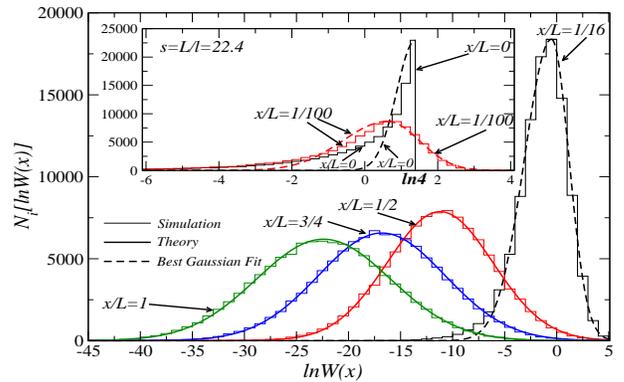}
}
\caption{
Evolution of the statistical distribution of $\ln {\cal W}(x)$ for
$x/L= 1.0, \; 3/4, \; 1/2$ and $1/16$, $x=L$ corresponding to $\ln T$.
The histograms are the results of computer simulations with $10^5$ realizations each,
and $kd=0.1$, $k\ell=178$.
All cases are 
in the localized regime, as $s=L/\ell = 22.4$.
The ordinate 
gives the number of events $N_i$ falling in box $i$ of the histogram.
The continuous curves are Gaussians with the parameters discussed in the text.
For $x$ not too close to zero, the agreement between the theoretically constructed Gaussians and the computed generated histograms is excellent.
}
\label{N(W(x))}
\end{figure}

We have not succeeded in finding the PDF of $\ln{\cal W}(x)$ analytically, but only 
numerically (Fig. \ref{N(W(x))}):
i) When $x$ is not too close to 0, e.g. for
$x/L=3/4 \; {\rm and} \; 1/2$, 
$\ln {\cal W}(x)$ has a 
normal PDF with the theoretical centroid and variance given in
Eqs. (\ref{<ln W>}) and (\ref{var(ln W)}).
When $x/L=1$, the PDF is normal, with 
$\langle \ln {\cal W}(L)\rangle = \langle \ln T\rangle$ and
${\rm var}[\ln {\cal W}(L)]_s={\rm var}[\ln T]_s$.
ii) For $x=0$, unitarity restricts $\ln {\cal W}(0) \le \ln 4$
(inset in Fig. \ref{N(W(x))}).
The PDF cannot be fitted by a truncated Gaussian:  
the dashed curve 
is the best ``half-Gaussian" fit (with the maximum at $\ln 4$) to the histogram.
iii) When $x \neq 0,L$, unitarity imposes 
no restriction on 
$\ln {\cal W}(x)$.
Close to the left end, 
$x/L \ll 1$, the PDF of
$\ln {\cal W}(x)$ admits non-zero values for 
$\ln {\cal W}(x) > \ln 4$. 
iv) For $x/L = 1/16$ (body of the figure) and $1/100$ (inset), the dashed curves show the best fit to the histograms 
by two ``half-Gaussians" 
on either side
of the maximum, 
using 
two different sets of parameters;
however, the left tail is longer than the Gaussian fit.

It is well known that the quantity $\ln T$ of Eqs. (\ref{<ln T>}) and (\ref{var(ln T)})
is self averaging [\onlinecite{lifshits_et_al-1988}],
whereas $T$ is not. 
Similarly, when $x\gg \ell$, one can show that $\ln {\cal W}(x)$ 
of Eqs. (\ref{<ln W>}) and (\ref{var(ln W)})
is self averaging, whereas ${\cal W}(x)$, studied in Ref. [\onlinecite{mello_genack_et_al_2015}], is not.
This is the main reason for studying $\ln {\cal W}(x)$ in the present paper
(see details in the SM of Ref. [\onlinecite{SM_1}]).

We have carried out microwave experiments to explore
the statistics of $\ln {\cal W}(x)$
inside random single-mode waveguides. Since
$\ln {\cal W}(x)$ self-averages, we are able to obtain sufficient
sampling to compare the measurements to theoretical
predictions in 100 random configurations.
Waves are launched
from one end of the waveguide and the signal is
detected by an antenna just above a slit along the length
of the waveguide.
The sample is composed of randomly positioned
elements contained within a
rectangular copper waveguide, with
width and height of 2.286 cm and 1.016 cm, giving a
cutoff frequency of 6.56 GHz.
The sample is made up of ceramic slabs with
dielectric constant $\epsilon = 15$, thickness of 0.66 cm
covering $93 \%$ of the waveguide cross section
and U-shaped Teflon elements which are
essentially air.
The elements in each
configuration are randomly selected with equal probability of being either a
dielectric or air layer.
The air layers may have thicknesses of
1.275, 2.550,
or 3.825 cm with equal probability.
The incident frequency ranges from 8.50 GHz
to 8.59 GHz in $400$ frequency steps.
The sample is of length $L$=60 cm.
The impact of absorption is removed by Fourier transforming the spectrum into the time
domain, multiplying by a factor $\exp({\Gamma_a t/2})$ and then transforming back into the frequency domain;
$\Gamma_a=0.011$ ${\rm ns}^{-1}$ is the decay rate of energy within the sample due to absorption and leakage through the slot along the sample length. It is obtained from the measurement of the linewidth in angular frequency units of the narrowest mode when copper reflectors are placed at the ends of the sample with only a small opening in the reflector on the LHS of the sample to admit energy from the source antenna. 
Absorbers are placed in the waveguide between the source antenna and the sample input and following the sample output to reduce reflection back into the sample.
\begin{figure}[t]
\includegraphics[width=9cm,height=6cm]{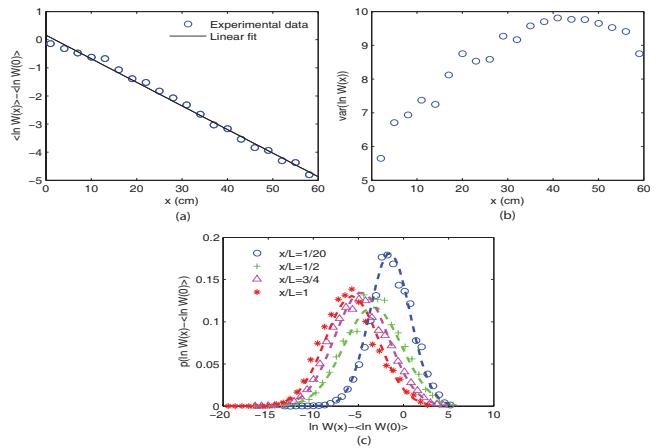}
\caption{
Results from microwave experiments for 
(a) $\langle \ln {\cal W}(x) \rangle$, 
(b) ${\rm var}(\ln {\cal W}(x))$. 
The solid line in (a) shows comparison to the theoretical prediction of a linear fall-off as in 
Eq. (\ref{<ln W>}).
(c) 
Experimental results for the PDF of $\ln {\cal W}(x)$ at different locations for $s=5$. 
Lines are Gaussian fit curves.
}
\label{expt}
\end{figure}
The experimental results
for $\langle \ln {\cal W}(x)\rangle_s$ shown in Fig. \ref{expt}a are well fit by the line $−5.06x/L + 0.02$.
This linear behavior is in agreement with 
the fit 
$s = L/\ell \approx 5$ and
$\ell \approx 12$ cm. 
Results for ${\rm var}[\ln {\cal W}(x)]_s$ are shown in 
Fig. \ref{expt}b: 
it increases linearly near
the beginning of the sample and bends as $x$
approaches the output boundary,
as in the theoretical result of
Fig. \ref{<lnW(x)>, var_lnW(x) theo. vs simuls.}b). 
However, ${\rm var}[\ln {\cal W}(0)]_s \sim 5$ is larger than the
predicted value of $\pi^2/3=3.29$. This is a consequence
of reflection by the source antenna. 
We find in 1D
simulations for a layered sample with an initial
layer with a high value of index of refraction $n_r$,
and hence of reflectivity, that the slope of 
$\langle \ln {\cal W}(x)\rangle_s$ is not affected by reflection from the
boundary, but ${\rm var}[\ln {\cal W}(0)]_s$ 
increases with $n_r$.
The PDF of $\ln {\cal W}(x)$ is shown in Fig. \ref{expt}c.
At the
beginning of the sample the
distribution is not symmetric; 
the fit shown in Fig \ref{expt}c
utilizes different Gaussian functions above and below the
peak value of the distribution.
However, for $x= L/2$ and $x = L$, 
the PDFs for
$\ln {\cal W}(x)$ are Gaussians as seen in Fig. \ref{expt}c. 
These results are consistent with features
seen in Fig. \ref{N(W(x))}.

In summary, we have used random-matrix theory to calculate the statistics of $\ln {\cal W}(x)$.
Since ${\cal W}(L)=T$, SPS corresponds to the particular case $x=L$, 
in the localized regime $s \gg 1$.
More generally, our analysis leads to the correction to SPS of Eq. 
(\ref{var(ln T)}).
Extending the MEA into the
interior of 1D samples provides a starting point for
analyzing the intensity inside systems supporting
several open channels.


The authors are indebted to B. Shapiro for valuable suggestions.
PAM and AZG wish to thank the Israel Institute of Technology (Haifa, Israel), where part of this work was started, for its hospitality.
PAM acknowledges support by DGAPA under contract No. IN109014,
and AZG the support of the National Science Foundation under grant 
No. 1609218 and the help of Noel Evans and Howard Rose for constructing the experimental assembly.







\end{document}



\title{SUPPLEMENTAL MATERIAL FOR THE PAPER: 
\vspace{5mm}

Single-Parameter Scaling and Maximum Entropy inside Disordered One-Dimensional Systems: Theory and Experiment
}

\author{Xujun Ma}
\thanks{These three authors contributed equally}
\address{Department of Physics, Queens College and the Graduate Center of the City University of New York, Flushing, NY, 11367 USA}

\author{Xiaojun Cheng}
\thanks{These three authors contributed equally}
\address{Department of Physics, Queens College and the Graduate Center of the City University of New York, Flushing, NY, 11367 USA}

\author{Miztli Y\'epez}
\thanks{These three authors contributed equally}
\affiliation{Departamento de F\'isica, Universidad Aut\'onoma Metropolitana-Iztapalapa, A. P. 55-534, 09340 M\'exico D. F., Mexico}

\author{Azriel Z. Genack}
\address{Department of Physics, Queens College and the Graduate Center of the City University of New York, Flushing, NY, 11367 USA}

\author{Pier A. Mello}
\address{Instituto de F\'isica, Universidad Nacional Aut\'onoma de M\'exico, Ap. Postal 20-364, 01000 M\'exico, D. F., Mexico}




\maketitle

\section{INSENSITIVITY of $\langle \ln {\cal W}(0) \rangle_s$ TO $L$}

We present, in the particular case $x = 0$, a simple demonstration of the independence of the expectation
value of the log of the intensity upon $L$
[Eq. (9) of the main text],
contrasting it with the $L$ dependence that appears when we do not take the logarithm.

In the DMPK approach one averages over phases 
(as we have done in the paper) 
and, as a result, $\langle r^n \rangle =0$, $r$ being the reflection {\em amplitude}.
For a 1D random potential this can be justified 
(see Ref. [34] of the main text)
by noting that the oscillations that still survive
as function of $kL$ are removed (to order $1/k\ell$) by averaging
$r$ over a small window 
[$\Delta (kL)/kL \sim 10^{-3}$ for the parameters used in Fig. 1 of the main text].

At $x=0$, the intensity is given by
\begin{equation}
{\cal W}(0) = (1+r)(1+r^{*}) \; .
\label{W(0)}
\end{equation}
Its expectation value is
\begin{eqnarray}
\langle {\cal W}(0) \rangle_s
&=& \langle  1+r+r^{*} + rr^{*} \rangle_s    \\
&=& 1+ \langle rr^{*} \rangle_s    \\
&=& 1+ \langle R \rangle_s 
= \left \{
\begin{array}{ccc}
 1, & {\rm for} & s = 0 \\
 & \vdots & \\
 2, & {\rm for} & s \to \infty
\end{array}
\right. ,
\label{<W(0)>}
\end{eqnarray}
showing clearly a dependence on $s=L/\ell$, 
as seen, for example, in Ref. [30] of the main text.

On the other hand, the expectation value of the log of the intensity of Eq. (\ref{W(0)}) can be calculated as
\begin{subequations}
\begin{eqnarray}
\langle \ln {\cal W}(0) \rangle_s
&=& \langle \ln (1+r) + \ln (1+r^{*})  \rangle_s \\
&=& \left\langle r - \frac{r^2}{2} + \frac{r^3}{3} + \cdots \right\rangle_s  
\nonumber \\
&&\hspace{5mm} + \left\langle r - \frac{r^2}{2} + \frac{r^3}{3} + \cdots \right\rangle_s^{*} \\
&=& 0,
\end{eqnarray}
\label{<lnW(0)>}
\end{subequations}
independent of $L/\ell$.

\section{the ``universal" value of ${\rm var}[\ln {\cal W}(0)]_s$}

Following a similar method, we evaluate in a simple way the ``universal" value of ${\rm var}[\ln {\cal W}(0)]$ in the localized regime.

We compute the second moment as
\begin{subequations}
\begin{eqnarray}
&& \langle [\ln {\cal W}(0)]^2 \rangle_s
= \langle [\ln (1+r) + \ln (1+r^{*}) ]^2  \rangle_s   
\\
&& = \left\langle 
\left(r - \frac{r^2}{2} + \frac{r^3}{3} + \cdots \right)^2 
\right.
\nonumber \\
 && 
+ \left(r^{*} - \frac{(r^{*})^2}{2} + \frac{(r^{*})^3}{3} + \cdots \right)^2 
\nonumber 
\\
&&
\left. \hspace{5mm} + 2\left(r - \frac{r^2}{2} + \frac{r^3}{3} + \cdots \right)
\left(r^{*} - \frac{(r^{*})^2}{2} + \frac{(r^{*})^3}{3} + \cdots \right)
\right\rangle_s 
\nonumber \\
&& = 2\left[\langle R \rangle_s + \frac{\langle R^2 \rangle_s}{2^2} + \cdots    \right] 
 \\
&& =  2 \sum_{n=1}^{\infty}\frac{\langle R^n \rangle_s}{n^2}
\label{<ln^2 W(0)> 1}
\end{eqnarray}
As $s\to \infty$, $\langle R^n \rangle_s \to 1$ and
\begin{eqnarray}
\langle [\ln {\cal W}(0)]^2 \rangle_s
&&\to 2 \sum_{n=1}^{\infty}\frac{1}{n^2}
\nonumber \\
&& = 2 \zeta(2) = 2\frac{\pi^2}{6} = \frac{\pi^2}{3}
\label{<ln^2 W(0)> 2}
\end{eqnarray}
\label{<ln^2 W(0)> 2}
\end{subequations}
This verifies, in an elementary way and for $x=0$, the more general result obtained in the paper for 
${\rm var}[\ln {\cal W}(x)]_s$, Eq. (12).
\vspace{1cm}


\section{$\langle \ln {\cal W}(x)\rangle_s$ and 
${\rm var}[\ln {\cal W}(x)]_s$ as functions of $s=L/l$}

The results shown in Fig. \ref{<lnW>,var[lnW]vs s} below complement
those of Fig. 1 of the main text: they present 
$\langle \ln {\cal W}(x)\rangle_s$ 
and ${\rm var}[\ln {\cal W}(x)]_s$ 
as functions of $s$ for fixed values of $x/L$.
They continue to the internal region of the sample the 
results for $\langle \ln T \rangle_s$ and 
${\rm var}[\ln T]_s$ at $x=L$ shown in the lowest curve in panel a) 
and the upper curve in panel b)
in Fig. \ref{<lnW>,var[lnW]vs s} of this supplemental material.
\begin{figure}[t]
\centerline{
\includegraphics[width=9cm,height=10cm]
{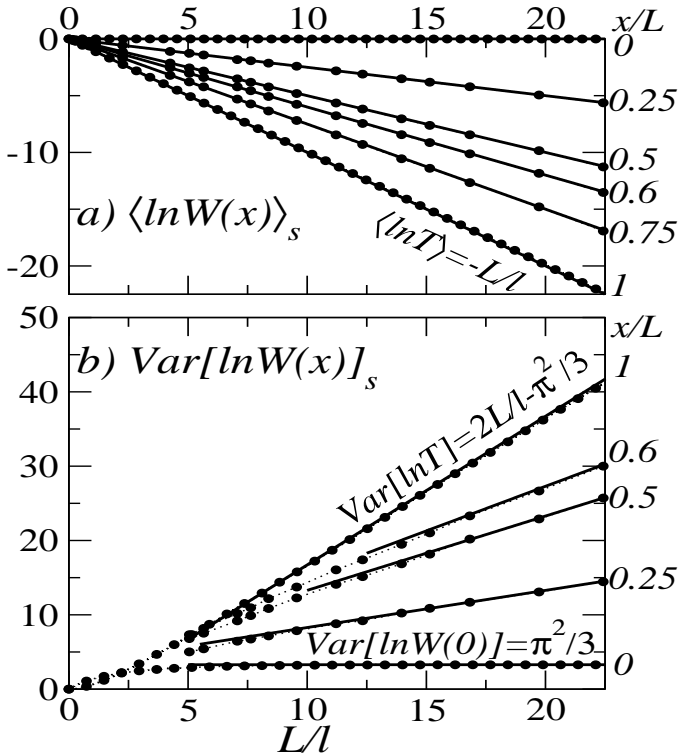}
}
\caption{
Theoretical results (full lines) and computer simulations
(various symbols and dotted lines)
for $\langle \ln {\cal W}(x)\rangle_s$ (panel a)) 
and  ${\rm var}[\ln {\cal W}(x)]_s$ (panel b)),
as functions of $s$ for fixed values of $x/L$.
Agreement is excellent:
in panel (a) in all cases;
in panel (b), when $s_2 \gg 1$. 
Simulations consist of $10^5$ realizations,
with $kd=0.1$ and $k\ell=178$.       
}
\label{<lnW>,var[lnW]vs s}
\end{figure}


\section{Self-averaging property of ${\cal W}(x)$ and $\ln {\cal W}(x)$}

From Eqs. (9) and (12) of the main text, the standard deviation (SD) 
$\sigma[\ln {\cal W}(x)]_s$ of the probability distribution of $\ln {\cal W}(x)$ is 
(for $s, s_2 \gg 1$)
\begin{equation}
\frac{\sigma [\ln {\cal W}(x)]_s}
{|\langle \ln {\cal W}(x) \rangle_s|} 
= \sqrt{2 \frac{\ell}{x}} \; .
\label{rel width lnW(x)}
\end{equation}
Provided the observation point $x$ is much farther away from the injection point than the MFP, i.e.,  $x \gg \ell$, the ratio (\ref{rel width lnW(x)}) becomes $\ll 1$
and we have a self-averaging quantity.

In contrast to the self averaging of $\ln {\cal W}(x)$, consider the intensity ${\cal W}(x)$.
As an example, at the midpoint $x=L/2$, 
the SD $\sigma \left({\cal W}(L/2)\right)$ of the probability distribution of ${\cal W}(L/2)$ is found to be
\begin{widetext}
\begin{equation}
\frac{\sigma \left({\cal W}(L/2)\right)}
{\left\langle {\cal W}(L/2) \right\rangle} \\
= \left[ \left\langle \frac{[1+3(\lambda_1+\lambda_2)+ 3(\lambda_1^2+\lambda_2^2)]
[1+ (\lambda_1+\lambda_2)+2\lambda_1\lambda_2]}
{[1+ (\lambda_1+ \lambda_2)]^3}  \right\rangle_{s/2, s/2}-1 \right]^{1/2} \; ,
\label{rel width lnW(L/2)}
\end{equation}
\end{widetext}
Here, $\langle {\cal W}(L/2)\rangle = 1$ (Ref. [30]) and $\lambda_1$, $\lambda_2$ are the radial variables described right after Eq. (5) of the main text. 
In contrast to the ratio in Eq. (\ref{rel width lnW(x)}), the ratio in Eq. (\ref{rel width lnW(L/2)})
increases with sample length, since it contains one more power of $\lambda_i$
in the numerator than in the denominator.

These results have important consequences for the ``noise" observed in averages performed over a finite number ${\cal N}$ of configurations of disorder; 
these averages are indicated by a bar over the quantity in what follows.
We write
\begin{equation}
\overline{\ln {\cal W}(x)}
\equiv \frac{1}{{\cal N}}
\sum_{i=1}^{\cal N}
[\ln {\cal W}(x)]_{{\rm i-th \; config.}},
\end{equation}
whereas $\ln {\cal W}(x)$ would correspond to ${\cal N}=1$.

The fluctuation of $\overline{\ln {\cal W}(x)}$
and of $\overline{{\cal W}(x)}$
over ${\cal N}$ configurations, relative to the centroid, are given by
\begin{subequations}
\begin{eqnarray}
\frac{{\sigma} \; \left[\overline{\ln {\cal W}(x)}\right]_s}
{\left|\langle \ln {\cal W}(x) \rangle_s\right|}
&=& \frac{1}{\sqrt{{\cal N}}}
\frac{{\sigma} \; \left[\ln {\cal W}(x)\right]_s}
{\left|\langle \ln {\cal W}(x) \rangle_s \right|} 
\label{noise lnW}  \\
\frac{{\sigma} \; \left[\overline{{\cal W}(x)} \right]_s}
{\langle {\cal W}(x) \rangle_s}
&=& \frac{1}{\sqrt{{\cal N}}}\frac{{\sigma} \; \left[{\cal W}(x)\right]_s}
{\langle {\cal W}(x) \rangle_s} 
\label{noise W}
\end{eqnarray}
\label{noise}
\end{subequations}

The second factor on the RHS of Eqs. (\ref{noise}) is the same as in Eqs. (\ref{rel width lnW(x)}), 
(\ref{rel width lnW(L/2)}) and contains
(theoretical) quantities evaluated over an unlimited number of configurations.
While the ``noise" in Eq. (\ref{noise lnW}) is negligible
(see Fig. 1a in the main text),
it is not so in Eq. (\ref{noise W}).
This is illustrated in Fig. \ref{barW,rms(barW)} of the present supplement,
which shows $\overline{{\cal W}(x)}$ in the two upper panels, for two values of $s$.
The size of the noise observed visually in the two upper panels is consistent with the result of Eq. (\ref{noise W})
obtained from $\sigma\left({\cal W}(x)\right)$ computed in the simulation
and shown in the two lower panels.

We can also give a rough estimate of ${\rm var}\left({\cal W}(L/2)\right)$ from the analytical result (\ref{rel width lnW(L/2)}).
As already noted, the first term in (\ref{rel width lnW(L/2)}) contains one more power of $\lambda_i$ in the numerator than in the denominator; thus the variance will grow exponentially with $s$. 
We can estimate 
${\rm var}\left({\cal W}(L/2)\right) \lesssim (3/2) {\rm e}^s \approx 2.2 \times 10^6$, for $s=14.2$, while the variance extracted from the simulation is 
$\approx 0.25 \times 10^6$.
\begin{figure}[b]
\centerline{
\includegraphics[width=9cm,height=6cm]
{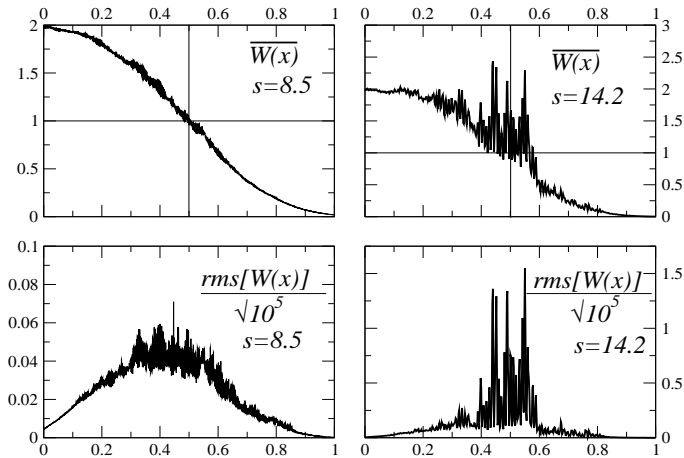}
}
\caption{
The two upper panels show computer simulations for $\overline{{\cal W}(x)}$ for a collection of $10^5$
configurations of disorder, for two lengths of the samples: $s=8.5$ and $s=14.2$.
It is evident how the ``noise" increases with increasing length.
The size of the noise shown in the two lower panels is computed as
the rms of $\overline{{\cal W}(x)}$ of the computer simulation data, and is consistent with the size of the noise visually observed in the two upper panels.
All the other parameters used in the simulations are the same as those in Fig. 1 of the main text.
The vertical and horizontal lines in the upper panels cross at the value 
$\langle {\cal W}(L/2)\rangle =1$ attained at the mid point $x=L/2$.
}
\label{barW,rms(barW)}
\end{figure}